\documentclass[12pt]{article}
 \usepackage{a4,epsfig,latexsym,amssymb,multirow,rotating,cite}

 \newcommand{\ts}{\textstyle}

 \newcommand{\half}{{\textstyle \frac{1}{2}} } 
 \newcommand{\third}{{\textstyle \frac{1}{3}} }

 \newcommand{\vma}{\delta{\mu}_a} 
 \newcommand{\vmb}{\delta{\mu}_b} 
 \newcommand{\vmq}{\delta{\mu}_q} 

 \newcommand{\mbar}{\overline{m}}

 \newcommand{\dmu}{\delta{m}_u} 
 \newcommand{\dmd}{\delta{m}_d} 
 \newcommand{\dms}{\delta{m}_s}

 \title{
\vspace{-3.25cm}
\flushright{\small ADP-15-30/T932} \\
\vspace{-0.35cm}
{\small DESY 15-158} \\
\vspace{-0.35cm}
{\small Edinburgh 2015/20} \\
\vspace{-0.35cm}
{\small Liverpool LTH 1055} \\
\vspace{-0.35cm}
{\small \today}  \\
\vspace{0.5cm}
 \centering{\Large \bf QED effects in the pseudoscalar meson sector}}

 \author{\large 
 R.~Horsley$^a$, Y.~Nakamura$^b$, H.~Perlt$^c$, D.~Pleiter$^d$,\\
 P.~E.~L.~ Rakow$^e$,
 G.~Schierholz$^f$, A.~Schiller$^c$, R.~Stokes$^g$, \\
 H.~St\"uben$^h$,
 R.~D.~Young$^g$ and J.~M.~Zanotti$^g$ \\[1em]
   -- QCDSF-UKQCD Collaboration -- \\[1em]
        \small $^a$ School of Physics and Astronomy,
               University of Edinburgh, \\[-0.5em]
        \small Edinburgh EH9 3FD, UK \\[0.25em]
        \small $^b$ RIKEN Advanced Institute for
               Computational Science, \\[-0.5em]
        \small Kobe, Hyogo 650-0047, Japan \\[0.25em]
        \small $^c$ Institut f\"ur Theoretische Physik,
               Universit\"at Leipzig, \\[-0.5em]
        \small 04109 Leipzig, Germany \\[0.25em]
        \small $^d$ J\"ulich Supercomputer Centre,
               Forschungszentrum J\"ulich, \\[-0.5em]
        \small 52425 J\"ulich, Germany, \\[-0.5em]
        \small Institut f\"ur Theoretische Physik,
               Universit\"at Regensburg, \\[-0.5em]
        \small 93040 Regensburg, Germany \\[0.25em]
        \small $^e$ Theoretical Physics Division,
               Department of Mathematical Sciences, \\[-0.5em]
        \small University of Liverpool,
               Liverpool L69 3BX, UK \\[0.25em]
        \small $^f$ Deutsches Elektronen-Synchrotron DESY, \\[-0.5em]
        \small 22603 Hamburg, Germany \\[0.25em]
        \small $^g$ CSSM, Department of Physics,
               University of Adelaide, \\[-0.5em]
        \small Adelaide SA 5005, Australia \\[0.25em]
        \small $^h$ Regionales Rechenzentrum, Universit\"at Hamburg, \\[-0.5em]
        \small 20146 Hamburg, Germany 
 }      

\begin{document}

 \maketitle


 \begin{abstract} 
 In this paper 
 we present results on the pseudoscalar meson masses from
 a fully dynamical simulation of QCD+QED, concentrating
 particularly on violations of isospin symmetry. 
 We calculate the $\pi^+$-$\pi^0$ splitting 
 and also look at other isospin violating mass differences. 
 We have presented results for these isospin splittings
 in~\cite{Horsley:2015}. In this paper we give more details
 of the techniques employed, discussing in particular the 
 question of how much of the symmetry violation is due to 
 QCD, arising from the different masses of the $u$ and $d$ 
 quarks, and how much is due to QED, arising from the 
 different charges of the quarks.  This decomposition is not unique, 
 it depends on the renormalisation scheme and scale.
 We suggest a renormalisation scheme in which Dashen's 
 theorem for neutral mesons holds, so that the electromagnetic
 self-energies of the neutral mesons are zero, and discuss
 how the self-energies change when we transform to a scheme
 such as $\overline{MS}$, in which Dashen's theorem for
 neutral mesons is violated. 
 \end{abstract} 

 \section{Introduction} 

 Lattice calculations of the hadronic spectrum are now reaching 
 a precision where it is essential to resolve  the influence of 
 isospin breaking effects. These have 
 two sources, a QCD effect arising from the fact that the $u$ and $d$
 quarks have different masses, and an electromagnetic effect due to 
 the $u$ and $d$ having different electric charges. The two effects 
 are comparable in magnitude, so a reliable calculation of isospin
 breaking requires simulating both the gluon and photon gauge fields. 

  Lattice studies of electromagnetic effects in the pions go
 back to~\cite{Duncan}. In recent years the interest in QCD+QED 
 has grown, and the pace of work
 accelerated~\cite{Blum,
  Aoki:2012st,deDivitiis:2013xla,Borsanyi:2013lga,Zhou:2014gga,
  Endres:2015gda, Borsanyi:2014jba}. 

  We are carrying out simulations in QCD+QED~\cite{Horsley:2015}. 
 Both gauge theories are fully dynamical, 
 so that the electrical charges of sea-quark loops are included 
 via the fermion determinants. We use a non-compact action 
 for the photon field.
  The calculations are carried out with three clover-like quarks. 
  Details of the lattice action will be given in section~\ref{lattdetail}, 
 and can be found in~\cite{Horsley:2015,Mainz2013}. 

   In the real world, with $\alpha_{EM}=1/137$, electromagnetic effects
 on masses are at the 1\% level, or smaller. This would make them 
 hard to measure on the lattice. Therefore 
 we simulate with a QED coupling stronger than in real world,
 so that we can see effects easily,
 and then scale back to physical $\alpha_{EM}$.  The simulations
 are carried out with 
 $ \beta_{QED} = 0.8$, equivalent to  $e^2 = 1.25,\ 
  \alpha_{EM} = e^2/(4\pi) \approx 0.10\,.$ 
 We will see that this is a good choice, 
 electromagnetic signals are clearly visible, much larger than 
 our statistical errors, but we are also in a region 
 where they still scale linearly in $e^2$, and we do not need to 
 consider higher-order terms. 

   We generate configurations with dynamical $u,d$ and $s$ quarks, 
 and then increase our data range by carrying out partially quenched 
 calculations, with valence  $u,d,s$ quarks having different masses from 
 the quarks used in the generation of the configurations.
 In addition to the  $u,d,s$ quarks, 
 we also introduce a fictitious $n$ quark, an extra flavour 
 with electrical charge zero. The $n$ quark is particularly useful 
 for checking that we are in the region where electromagnetic 
 effects are still linearly proportional to $e^2$. 

  In this work we present results on the pseudoscalar mesons. 
 Our meson propagators are calculated from connected graphs only.
 Because we have no fermion-line disconnected graphs, 
 the $u \bar u, d \bar d, s \bar s$ and $ n \bar n$ states do
 not mix, so we can measure $M^2(u \bar u), M^2(d \bar d)$ and 
 $M^2(s \bar s)$. In the real world, these states do not exist, 
 they mix strongly to form the $\pi^0, \eta$ and $\eta^\prime$. 
 Disconnected graphs are responsible for the large mass of the
  $\eta^\prime$, but will have very little effect on the mass
 of the $\pi^0$. 
 In this work we do not consider the $\eta$ and  $\eta^\prime$
 further, but we will need 
 a mass for the $\pi^0$, with wave-function 
 proportional to $( u \bar u - d \bar d )/\sqrt{2}$. 
 We use the relation 
 \begin{equation} 
   M^2_{\pi^0} \approx \half \left[M^2(u \bar u) + M^2(d \bar d) \right]  
 \end{equation} 
 which is a very good approximation, with corrections proportional to
 the small quantity $(m_d-m_u)^2$~\cite{groupy}. 
 This issue does not arise for the flavour non-diagonal mesons, 
 $\pi^+, K^0, K^+,$ which have no disconnected contribution. 

   In the first part of this paper, sections~\ref{extrap} 
 to~\ref{scheme},
 we discuss theoretical
 questions. First we describe how our constant singlet mass
 procedure~\cite{strange1,groupy} can be applied to QCD+QED. 
 We derive a mass formula for pseudoscalar mesons in this framework. 
 This is all that is needed to calculate physical mass splittings, 
 in particular the $\pi^+$-$\pi^0$ splitting.  
 It also gives us the lattice masses for the $u, d, s$ quarks at
 the physical point, needed to predict mass splittings in the baryons.
 A particularly delicate number is the mass difference $m_u-m_d$
 (or $m_u/m_d$ mass ratio), which is difficult to extract 
 reliably from a pure QCD simulation, and is much better defined
 in QCD+QED simulations.

    We also want to dissect the meson mass into a QCD part and a QED part,
 to find the electromagnetic $\epsilon$ parameters, which express 
 the electromagnetic contributions to the meson masses ~\cite{FLAG}.
 We find that
 there are theoretical subtleties in this separation, leading to 
 scheme and scale dependence in the result. 

    The total energy-momentum tensor is invariant under renormalisation, 
 and so the total mass of any hadron is independent of renormalisation
 scheme and scale. However the individual contributions from quarks, 
 gluons and photons are not invariant, they all run as the energy
 scale increases. This is familiar in pure QCD; as the energy scale
 of Deep Inelastic Scattering rises, the momentum fraction carried
 by quarks decreases, while the momentum fraction carried by gluons
 increases~\cite{AltarelliParisi}. The physical picture behind this
 effect is well known~\cite{textbooks}.
 As $Q^2$ rises the proton is probed with 
 improved spatial resolution. A parton perceived as a  single 
 quark in a low-$Q^2$ measurement is resolved into multiple 
 partons at higher $Q^2$, with most of the new partons being gluons. 
 
   We should expect a similar effect in QCD+QED, with improved
 spatial resolution revealing more photons, causing a running
 of energy from quarks to photons, in parallel with the running from 
 quarks to gluons seen in QCD alone. 

    In QCD+QED, each hadron will be surrounded by a 
  photon cloud. As in pure QED, the total energy in the cloud will 
  be ultra-violet divergent. 
    Crudely, we can think of two components of the cloud.
   Firstly, there are
 short wave-length photons, with wave-lengths small compared 
  with a hadron radius. These can be associated with particular
  quarks. If we look at the hadron with some finite resolution
  the photons with wavelengths shorter than this resolution are
  incorporated into the quark masses as self energies. 
  Secondly, there will be 
   longer wave-lengths photons, which can't be associated with particular 
  quarks. These photons must be thought of as the photon cloud 
  of the hadron as a whole, these are the photons that we 
  include when we talk of the electromagnetic contribution to the 
  hadron mass. We expect to see many more really long wave-length
  photons (large compared to the hadron radius)  around a 
  charged hadron than around a neutral hadron. 
 
     Clearly, in this picture, the value we get for the electromagnetic
  contribution to the hadron energy is going to depend on our 
  resolution, i.e. on the scheme and scale that we use for 
  renormalising QED. 
 
     In the final part, section~\ref{results}, we summarise our 
  lattice results for the $\pi^+$-$\pi^0$ splitting and for 
  the scheme-dependent $\epsilon$ parameters, which parameterise
  the electromagnetic part of the meson masses. 
 
     We have already published an investigation into the QCD
  isospin breaking arising from $m_d-m_u$ alone in~\cite{isospin}, 
  and the first results of our QCD+QED program in~\cite{Horsley:2015}, 
  which we discuss at greater length here. 
 
 \section{Extrapolation Strategy \label{extrap}} 

   In pure QCD we found that there are significant advantages
 in expanding about a symmetric point with $m_u = m_d = m_s
 = \overline{m}\;$~\cite{strange1,groupy}. In particular, 
 this approach simplifies the extrapolation to the physical point, 
 and it decreases the errors due to partial quenching. 
  We want to follow a similar approach with QED added,
 even though the symmetry group is smaller (the $u$ quark
 is always different from the other two flavours because
 of its different charge).

 First we find a symmetric point, with all three quark masses
 equal, chosen so that the average quark mass, 
 \begin{equation} 
 \mbar \equiv \third \left(m_u + m_d + m_s \right) \;, \label{mbar}
 \end{equation} 
 has its physical value. To do this, we have defined our
 symmetric point in terms of the masses of neutral pseudoscalar mesons
 \begin{equation}
 M^2( u \bar u) = M^2( d \bar d) = M^2( s \bar s) = M^2( n \bar n)
 = X_\pi^2 \;.
 \label{symdef}
 \end{equation}
   Here $X_\pi$ is an average pseudoscalar mass, defined by
 \begin{equation}
 X_\pi^2 = \third \left[ 2 (M_K^\star)^2 + (M_\pi^\star)^2\right]
 \label{Xpidef} 
 \end{equation}
 where $\star$ denotes the real-world physical value of a mass.
 The $n$ is a fictitious electrically neutral quark flavour.
 We have not included disconnected diagrams, so the different
 neutral mesons of~(\ref{symdef}) do not mix. 

  We also define the critical $\kappa^c_q$ for each flavour as the
 place where the corresponding neutral meson is massless
 \footnote{The critical $\kappa$ defined in eq.~(\ref{kcdef})
 is the  critical $\kappa$ in the $m_u + m_d + m_s
 = const$ surface, i.e. if $m_u = 0$, we must have
 $m_d + m_s = 3 \overline m$. The $\kappa^c$ for the chiral point with
 all three quarks massless will be different.} 
 \begin{equation}
 M^2(q \bar q) = 0 \quad  \Leftrightarrow  \quad m_q = 0 \; .
 \label{kcdef} 
 \end{equation}
 Chiral symmetry can be used to argue that neutral mesons
 are better than charged ones for defining the massless
 point~\cite{Das}.

 We then make a Taylor expansion about this point, using the 
 distance from $\mbar$ as our parameter to specify the bare quark
 masses
 \begin{eqnarray} 
 a\delta m_q &\equiv& a( m_q - \mbar)
 = \frac{1}{2 \kappa} - \frac{1}{2 \kappa_q^{sym}}\;,\\
 a\delta \mu_q &\equiv& a( \mu_q - \mbar)
 = \frac{1}{2 \kappa} - \frac{1}{2 \kappa_q^{sym}}\;, 
 \end{eqnarray} 
 where $m_q$ denotes the simulation quark mass (or sea quark mass), 
 while $\mu_q$ represents the masses of partially quenched valence
 quarks. Note that keeping the average quark mass constant,~(\ref{mbar}),
 implies the constraint
 \begin{equation} 
 \dmu + \dmd + \dms = 0 \;. 
 \end{equation} 

   In~\cite{groupy} 
   we wrote down the allowed expansion terms for pure QCD, 
 taking flavour blindness into account. QCD+QED
 works very much like pure QCD.  
 Since the charge matrix $Q$ is a traceless $3 \times 3$ 
 matrix, 
 \begin{equation} 
 Q = \pmatrix{ \textstyle +\, \frac{2}{3} & 0 & 0 \cr 
 0 & -\, \frac{1}{3} & 0 \cr 0 & 0 & -\, \frac{1}{3} } \;,
 \end{equation} 
 electric charge is an octet, so we can 
 build up polynomials in both charge and mass splitting
 in a way completely analogous to the pure QCD case.
   The main difference is that we can only have even powers of the
 charge, so the leading QED terms are 
  $\sim  e^2$, while the leading QCD terms are $\sim \delta m$.
 
  One very important point to note is that even when all three quarks
 have the same mass, we do not have full SU(3) symmetry. The different
 electric charge of the $u$ quark means that it is always distinguishable
 from the $d$ and $s$ quarks. 

 \section{Meson mass formula}

  From these considerations 
  we find the following expansion for the mass-squared of an 
 $a \bar b$ meson, incorporating both the QCD and electromagnetic
 terms   
 \begin{eqnarray}
 M^2(a \bar b) &=& M^2 + \alpha ( \vma + \vmb) +c (\dmu + \dmd + \dms) 
 \label{QEDmeson} 
  \\ &&{}
 + \beta_0 {\ts \frac{1}{6}} ( \dmu^2 + \dmd^2 + \dms^2)
 + \beta_1 ( \vma^2 + \vmb^2)
 + \beta_2 ( \vma - \vmb)^2  \nonumber
  \\ &&{}
 + \beta_0^{EM} (e_u^2 + e_d^2 + e_s^2)
  + \beta_1^{EM} (e_a^2 + e_b^2)
 + \beta_2^{EM} (e_a - e_b)^2  \nonumber
  \\ &&{}
 + \gamma_0^{EM} ( e_u^2 \dmu + e_d^2 \dmd + e_s^2 \dms )
  + \gamma_1^{EM} ( e_a^2 \vma + e_b^2 \vmb ) \nonumber
 \\ &&{}
 + \gamma_2^{EM} (e_a - e_b)^2 (\vma + \vmb )
 + \gamma_3^{EM} (e_a^2 - e_b^2) ( \vma - \vmb ) \nonumber
  \\ &&{}
 + \gamma_4^{EM} (e_u^2 +e_d^2 + e_s^2) (\vma + \vmb) \nonumber
  \\ &&{}
 +  \gamma_5^{EM} (e_a + e_b) (e_u \dmu + e_d \dmd + e_s \dms)
 \;.  \nonumber
 \end{eqnarray}  
 As well as the terms needed in the constant $\mbar$ surface we have
 also included the term $c (\dmu + \dmd + \dms)$, the leading term 
 describing displacement from the constant $\mbar$ surface. Including
 this term will be useful when we come to discuss renormalisation 
 and scheme dependence, it could also be used to make minor 
 adjustments in tuning. 

 The QCD terms have been derived in~\cite{groupy}. In particular, 
 we discussed the effect of chiral logarithms in section~V.C. of
 that paper. Briefly, since we are expanding about a point
 some distance away from all chiral singularities the chiral 
 logarithms do
 not spoil the expansion, but they do determine the behaviour of
 the series for large powers of $\delta m_q$, (see for example equation
 (78) of~\cite{groupy}).

 We will now
 discuss briefly the origins of the electromagnetic terms. 

    \subsection{Leading order terms}

 In what follows we use the following notation: 
 \begin{equation} 
 e^2 = 1/\beta_{QED}\;,  \qquad e_q = Q_q e 
 \end{equation} 
 where 
 \begin{equation} 
 Q_u = +\,{\ts  \frac{2}{3}\;,} \quad  Q_d = Q_s = -\, {\ts \frac{1}{3}}  \;. 
 \end{equation} 

 The leading order EM terms were written down in~\cite{Mainz2013},
 \begin{equation}
 M^2_{EM}(a \bar b) =
 \beta_0^{EM} (e_u^2 + e_d^2 + e_s^2)
  + \beta_1^{EM} (e_a^2 + e_b^2) + \beta_2^{EM} (e_a - e_b)^2 \;. 
 \end{equation}

 Upon examination of each of these terms in more detail, we observe
 that since all of our simulations have the same choice of sea quark
 charges, then even if we vary the sea quark masses, $(e_u^2 + e_d^2 +
 e_s^2)$ is a constant, and we can simply absorb this term into $M^2$
 of~(\ref{QEDmeson}).
 Hence, the $\beta_0^{EM}$ term just stands for the fact that $M^2$
 measured in QCD+QED might be different from $M^2$ measured in pure
 QCD.  As we have tuned our expansion point so that the pseudoscalars
 have the same symmetric-point mass as in pure QCD, the $\beta_0^{EM}$
 for the pseudoscalar mesons will be zero, but we will still have to
 allow $M^2$ for other particles to be different in QCD+QED than in
 pure QCD.

      Now consider~(\ref{QEDmeson}) at the symmetric point, for the 
 case of a flavour-diagonal meson, $a \bar a$. At the symmetric 
 point, nearly all terms vanish because $\delta m_q$ and $\delta \mu_q$
 are zero. In addition, the electromagnetic terms simplify
 because $e_b=e_a$. All we are left with is 
 \begin{equation} 
 M^2(a \bar a) = M^2 + \beta_0^{EM} (e_u^2 + e_d^2 + e_s^2)
  + 2 \beta_1^{EM} e_a^2 
 \label{symM2}
 \end{equation} 
 at the symmetric point. 
 However, since we have defined our symmetric point by~(\ref{symdef}), 
 equation~(\ref{symM2}) must give the same answer whether $e_a = 
 -\;\frac{1}{3}e, 0$ or $+\frac{2}{3}e$, so $\beta_1^{EM} $
 must be zero (because it would split the masses of the different
 mesons, according to the charge of their valence quarks).
 However, having  $\beta_1^{EM} =0$ for the pseudoscalar
 mesons does not mean that this term will also vanish for
 other mesons, for example the vector mesons. If we tune
 our masses so that the pseudoscalar $u \bar u$, $d \bar d$ and 
 $s \bar s$ all have the same mass, we would still expect to find that
 the vector $u \bar u$ meson would have a different mass from the 
 vector  $d \bar d$ and $s \bar s$, because there is no symmetry
 in QCD+QED which can relate the $u$ to the other two flavours. 

 Finally, we observe that the contribution from 
 $\beta_2^{EM}$ is zero for neutral mesons, $e_a=e_b$. However, this is
 the leading term contributing to the $\pi^+$-$\pi^0$ mass splitting, 
 so it is of considerable physical interest. 

 \subsection{Next Order} 

 Going beyond leading order, the following higher order terms of the
 form $e^2 \delta m_q$, $e^2 \delta \mu_q$ are possible:

 \begin{itemize} 

  \item{ Sea charge times sea mass, $\gamma_0^{EM}$ }

   After imposing the constraints that $\mbar$ is kept
 constant and $e_u+e_d+e_s=0$, there is only one completely
 symmetric sea-sea polynomial left,
 \begin{equation}
 e_u^2 \dmu + e_d^2 \dmd + e_s^2 \dms \;. 
 \end{equation}

 \item{ Valence charge times sea mass }

  At this order all polynomials of this type are killed by
 the $\mbar = const$ constraint.

 \item{ Valence charge times valence mass,
 $\gamma_1^{EM}, \gamma_2^{EM}, \gamma_3^{EM}$ }

   In this case there are three independent allowed terms.
 One convenient basis for the valence-valence terms is
 \begin{equation}
   e_a^2 \vma + e_b^2 \vmb\;,  \qquad
  (e_a - e_b)^2 (\vma + \vmb )\;, \qquad
  (e_a^2 - e_b^2) ( \vma - \vmb )\;, 
 \end{equation}
 though other choices are possible.

 \item{ Sea charge times valence mass, $\gamma_4^{EM}$ }

  The only polynomial of this type is
 \begin{equation}
 (e_u^2 + e_d^2 + e_s^2) ( \vma + \vmb )\;. 
 \end{equation}
 Since $(e_u^2 + e_d^2 + e_s^2)$ is held constant,
 this term can simply be absorbed into the parameter
 $\alpha$ of~(\ref{QEDmeson}).

 \item{ Mixed charge times sea mass, $\gamma_5^{EM}$ }

  At the symmetric point we can not have mixed charge
 terms (valence charge times sea charge), because
 such terms would be proportional to $(e_u+e_d+e_s)$ 
 which is zero. 
  However, away from the symmetric point
 \begin{equation}
 (e_a + e_b) (e_u \dmu + e_d \dmd + e_s \dms)
 \end{equation}
 is allowed. 

 \end{itemize} 

\begin{figure}[!htb]
   \begin{center}
 \epsfig{file=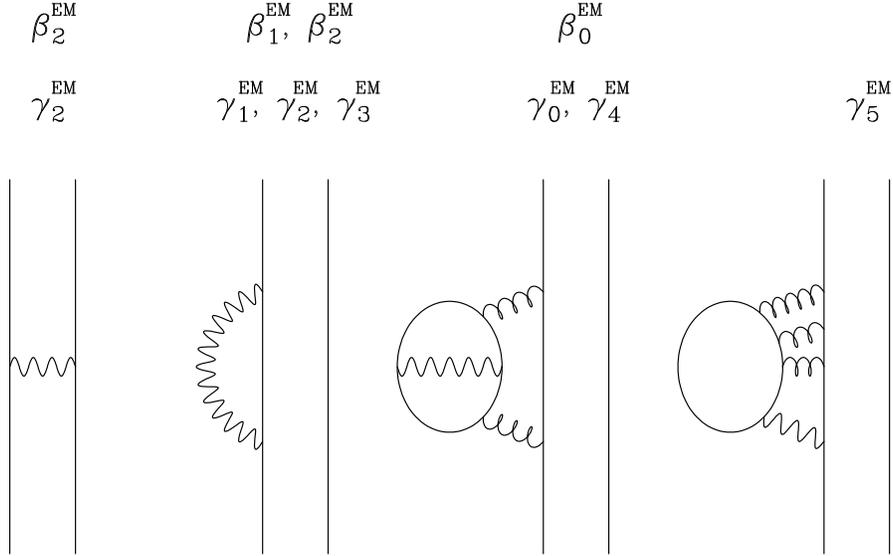,width=13cm}
 \caption{Examples of the Feynman diagrams contributing to each 
 of the electromagnetic coefficients in the meson mass 
 formula~(\ref{QEDmeson}). All the graphs have a single photon 
 (wavy line), and are all of $O(e^2)$ in the electromagnetic 
 coupling. However, some terms require multiple gluons (curly lines), 
 and so have higher order in the strong coupling $g^2$. 
 \label{Feyn} }
  \end{center} 
 \end{figure}

 We illustrate the different physical origins of these terms by 
 drawing examples of the Feynman diagrams contributing to each of 
 the electromagnetic coefficients in~(\ref{QEDmeson}), Fig.~\ref{Feyn}. 

 \section{Lattice setup \label{lattdetail}} 

   We are using the action
 \begin{equation}
 S = S_G + S_A + S_F^u + S_F^d + S_F^s \;.
 \end{equation}
 Here $S_G$ is the tree-level Symanzik improved SU(3) gauge action,
 and $S_A$ is the noncompact U(1) gauge action of the photon,
 \begin{equation}
 S_A = \half \beta_{QED} \sum_{x, \mu < \nu}
 \left[ A_\mu(x) + A_\nu(x + \hat\mu) -
 A_\mu(x + \hat\nu) - A_\nu(x) \right]^2 \;.
 \end{equation}
 The fermion action for flavour $q$ is
 \begin{eqnarray}
 S_F^q &=& \sum_x {\Bigg\{} \half \sum_\mu \left[
 \overline{q}(x) (\gamma_\mu -1) e^{-i Q_q A_\mu(x) }
 \tilde {U}_\mu(x) q( x + \hat\mu) \right.
  \nonumber \\
  &&{}\qquad \left. - \overline{q}(x)
 (\gamma_\mu +1) e^{ i Q_q A_\mu(x-\hat\mu) }
  \tilde {U}^\dagger_\mu(x-\hat\mu) q(x-\hat\mu) \right] \nonumber \\
  &&{}\qquad +\; \frac{1}{2 \kappa_q} \overline{q}(x) q(x)
  - {\ts \frac{1}{4} } c_{SW} \sum_{\mu,\nu}
 \overline{q}(x) \sigma_{\mu \nu} F_{\mu \nu}(x) q(x) \Bigg\} \;,
 \end{eqnarray}
 where $  \tilde {U}_\mu$ is a singly iterated stout link.
 We use the clover coefficient $c_{SW}$ with the value computed
 non-perturbatively in pure QCD,~\cite{csw}. We do not include a clover
 term for the electromagnetic field.
 We simulate this action using the Rational Hybrid
 Monte Carlo (RHMC) algorithm~\cite{RHMC}.
 
  One issue that arises in the simulation of QED is the treatment 
 of constant electromagnetic background fields. In simulations 
 where the electromagnetic field does not couple to the quark 
 determinant these are electromagnetic zero modes, and so need to 
 be handled with particular care. In this
 simulation the sea quarks are coupled to the electromagnetic field, 
 and so the action does depend on the background field. However 
 we do still need to give special treatment to these modes.  
  We handle constant background fields by adding or subtracting
  multiples of $6 \pi /(e L_\mu) $ until the background field is
  in the range 
  \begin{equation} 
  -3 \pi < e B_\mu L_\mu \le 3 \pi 
  \end{equation} 
  This is the mildest way to keep the background fields under control
  \cite{Gockeler:1991bu}. This procedure leaves fermion determinants
  unchanged for particles with charges a multiple of $e/3$. It also
  leaves Polyakov loops unchanged (again, for charges in units of
  $e/3$).  We are investigating the evolution of these background
  fields in our simulations, and considering what effect they have on
  finite size effects. We plan to report on these studies in a future
  paper.

  We have carried out simulations on three lattice volumes, 
 $24^3 \times 48, 32^3 \times 64$ and $48^3 \times 96$. 
  The $24^3 \times 48$ calculations show clear signs of finite
 size effects. The differences between $32^3 \times 64$ and
 $48^3 \times 96$ are quite small, leading us to believe
 that finite size effects on our largest volume are under control. 
 In this paper we present results from the two largest volumes, 
 which usually are in close agreement. In the few cases where there 
 is a difference, we would favour the results from the largest 
 volume,  $48^3 \times 96$. 
 
 \section{Critical $\kappa$}

    After several tuning runs we have been carrying out our main
 simulations at the point
 \begin{eqnarray}
 \beta_{QCD} = 5.50  \;, & \quad & \beta_{QED} = 0.8 \;,  \\
 \kappa_u = 0.124362\;, &&  \kappa_d = \kappa_s =0.121713  \nonumber
 \end{eqnarray}
 which lies very close to the ideal symmetric 
 point defined in~(\ref{symdef}) 
 (but with a much stronger QED coupling than the real world,
 $\alpha_{QED} = 0.099472\cdots $, instead of the true value $1/137$).
 At this point the $\delta m_q$ from the sea quark masses are all zero,
 but we can still learn abut the meson masses by varying the partially
 quenched valence quark masses, $\delta \mu_q$.  

\begin{figure}[!htb]
   \begin{center}
 \epsfig{file=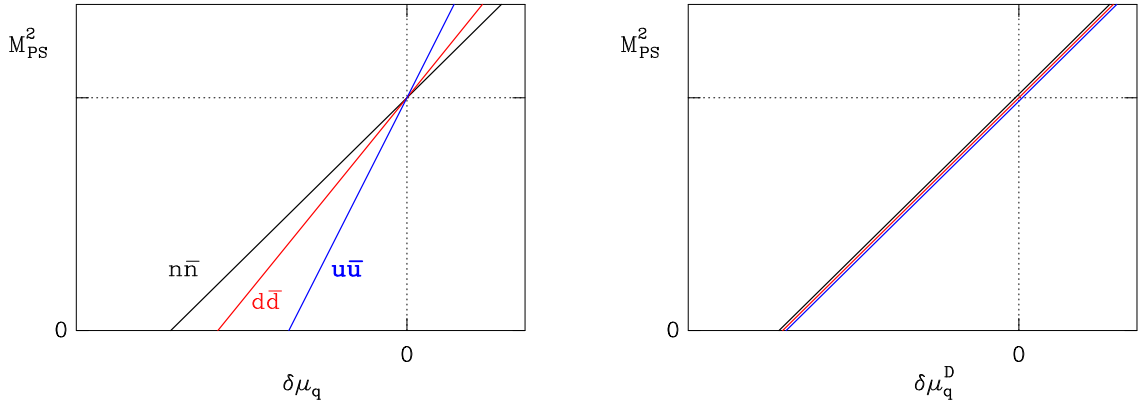,width=15cm}
 \caption{Sketch illustrating the transformation from bare masses
 (left panel) to Dashen scheme masses (right panel). In the left
 panel all the flavour diagonal mesons have the same mass at 
 the symmetric point ($\vmq=0$), but have different critical
 points ($M^2_{\rm PS}=0$). 
 In the Dashen scheme (right panel) we rescale the masses horizontally, 
 so that all the critical points are the same. The different 
 mesons now all depend on $\vmq^D$  in the same way. 
 \label{pisketch} }
  \end{center} 
 \end{figure}

    The flavour dependence of the meson masses is more complicated
 in QCD+QED than in pure QCD. We illustrate some of these 
 differences in the sketch Fig.~\ref{pisketch}, showing the way that
 the flavour-diagonal mesons depend on the quark mass. 
 As well as the physical charge $+\frac{2}{3}$ and $-\frac{1}{3}$ 
 quarks, we also have a fictional charge 0 quark. 
 In QCD+QED we still have the relationship $M^2(q \bar q) \propto m_q$
 for flavour-diagonal (neutral) mesons, but the gradients of the
 $u \bar u, d \bar d, n \bar n$ mesons differ. So, in contrast to
 pure QCD, equal meson mass at the symmetric point 
 no longer means equal bare quark mass. 
 The bare mass at the symmetric point depends on the quark charge.
 This situation is illustrated in the left panel of Fig.~\ref{pisketch}, 
 (though the differences between the flavours has been exaggerated 
 for clarity).

   We rescale (renormalise) the quark masses to remove this effect, 
 making the renormalised quark masses at the symmetric point
 equal. The situation after renormalising in this way is illustrated 
 in the right panel of Fig.~\ref{pisketch}. All the 
 flavour-diagonal mesons, $n \bar n, d \bar d, s \bar s$ and
 $u \bar u$ now line up, depending in the same way on the 
 new mass $\mu^D$, which we call the ``Dashen scheme" mass, 
 for reasons which should become clear later  
\footnote{Here,
 to introduce the idea, we just make a simple multiplicative 
 renormalisation. In fact, the mass renormalisation matrix is
 not diagonal, there are also terms which mix flavours. We will
 include these additional terms in section~\ref{Dashq}.}.
 We will see that using this quark mass also simplifies 
 the behaviour of the mixed flavour mesons, and helps us
 understand the splitting of a hadron mass into a QCD part
 and an electromagnetic part. 

   One way to interpret the behaviour in
 Fig.~\ref{pisketch} is to consider a $u$ and 
 $d$ quark with the same bare lattice mass. 
 Since the magnitude of the charge of the $u$ quark is twice as large
 as that of the $d$ quark, it will acquire a larger self-energy due to
 the surrounding photon cloud and hence it will be physically more
 massive, which is why the mass of the $u \bar u$ meson rises more
 steeply than the $d \bar d$ meson, when plotted against bare mass. By
 instead plotting against the Dashen mass, we have effectively added
 the extra mass of the photon cloud to the quark mass. Two quarks with
 the same Dashen mass are physically similar in mass, and so they form
 mesons of the same mass, as seen in the right-hand panel of
 Fig.~\ref{pisketch}.

\begin{figure}[!htb]  \begin{center}
 \epsfig{file=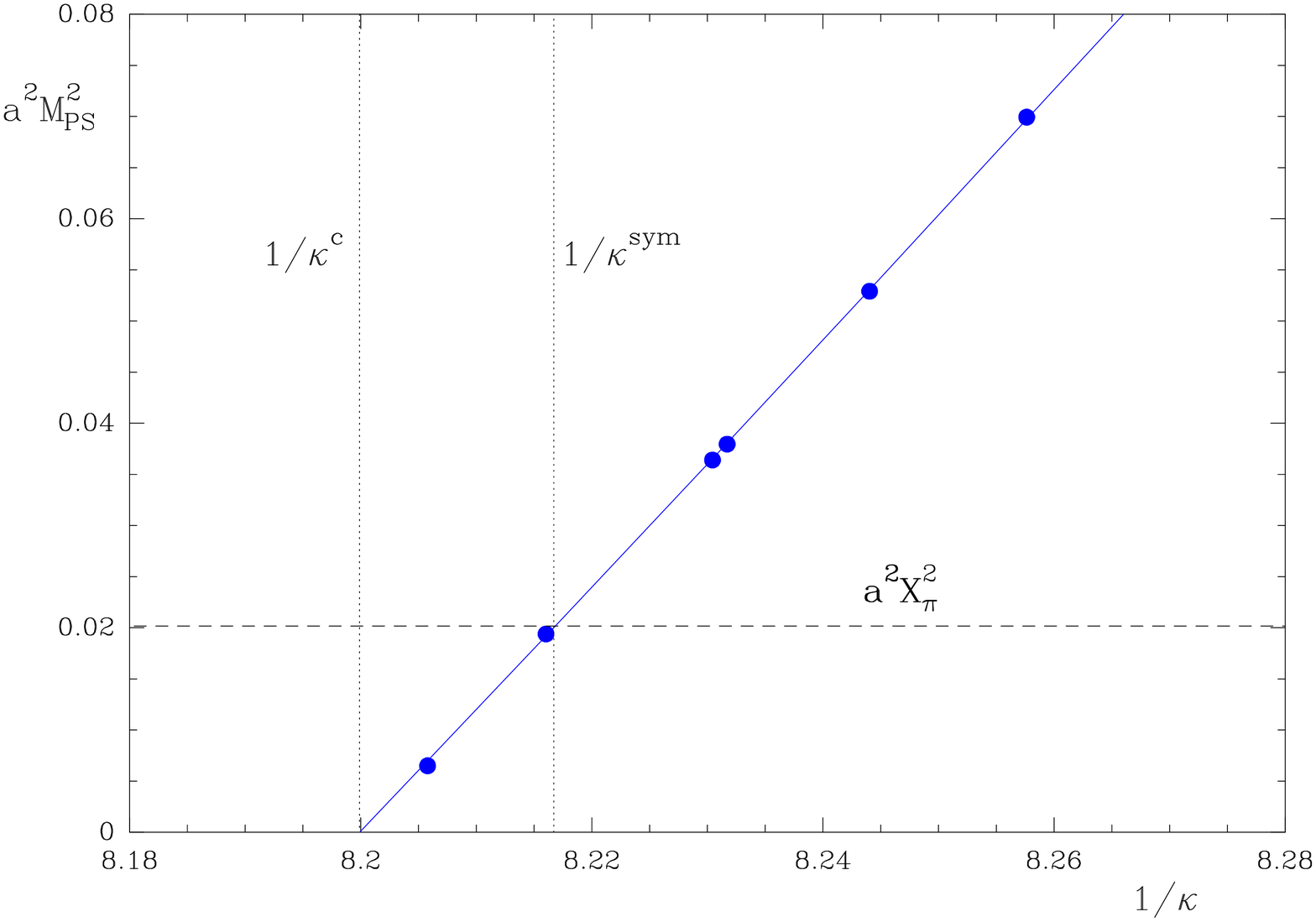,width=11cm}
 \caption{ Determination of $\kappa^c$ and $\kappa^{sym}$ 
 for the $d$ quark. $\kappa^c$ is defined from the point where the
 $d \bar d$ meson mass extrapolates to zero,~(\ref{kcdef}), while
 $\kappa^{sym}$ is defined by the point where the fit line
 crosses $M_{PS}^2 = X_\pi^2$,~(\ref{symdef}).
 \label{kkdd} }
 \end{center}
 \end{figure}

 Applying these ideas to our simulations, in Fig.~\ref{kkdd} we show
 how the symmetric $\kappa^{sym}$ and critical $\kappa^c$ are
 determined, using the $d \bar d$ meson as an example. $\kappa^c$ is
 defined from the point where the partially-quenched meson mass
 extrapolates to zero,~(\ref{kcdef}), while $\kappa^{sym}$ is defined
 by the point where the fit line crosses $M_{PS}^2 =
 X_\pi^2$,~(\ref{symdef}).

\begin{figure}[!htb]  \begin{center}
 \epsfig{file=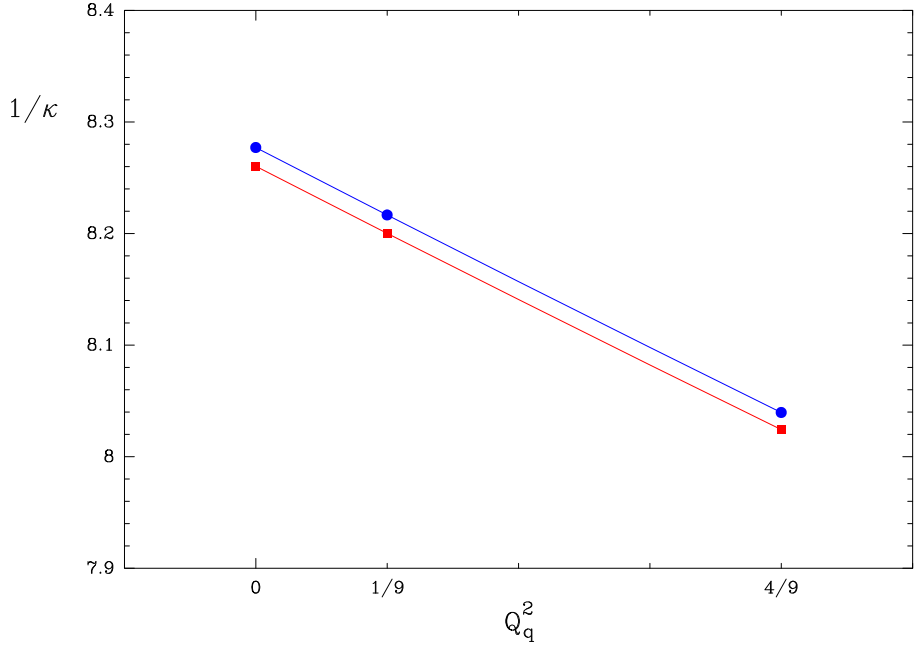,width=12cm}
 \caption{  $1/\kappa^c$ (red squares) and $1/\kappa^{sym}$
 (blue circles) plotted
 against quark charge squared, $Q_q^2$.
 \label{kaplot} }
 \end{center}
 \end{figure}

 We repeat this procedure for the $u$ and $n$ quarks and plot the resulting
 $1/\kappa^c$ and $1/\kappa^{sym}$ values as a function of the square
 of the quark charges, $Q_q^2$, in Fig.~\ref{kaplot} 
 Here we clearly see that in both cases $1/\kappa$ depends linearly on $Q_q^2$. 

\begin{figure}[!htb]  \begin{center}
 \epsfig{file=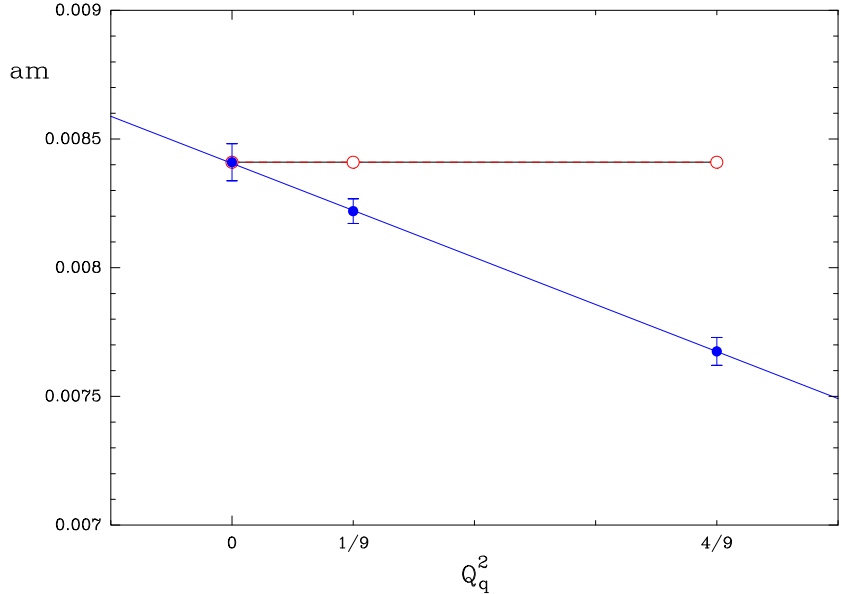, width=11cm} 
 \end{center} 
 \caption{The  bare mass at the symmetric
 point, $ a m_q^{sym}$, as a function of quark charge. 
 We see that the bare mass is not constant, there 
 is about a 10\% difference between the neutral $n$ quark 
 and the $u$ quark. The open red circles show the quark masses
 after renormalising to remove this charge dependence. 
 \label{bareplot} } 
 \end{figure}

 Despite appearances, the two lines are not quite parallel. 
 In Fig.~\ref{bareplot} we plot the bare mass at the symmetric point,
 \begin{equation} 
 a m^{sym}_q = \frac{1}{2 \kappa_q^{sym}} - \frac{1}{2 \kappa_q^c}  \;. 
 \end{equation} 
 $\kappa_q^c$ for each flavour is defined as the point at which 
 the flavour-diagonal $q \bar q$ meson becomes massless. We see
 that our data show the behaviour shown in the left-hand panel
 of Fig.~\ref{pisketch}, with each meson reaching the axis at a different
 point. 

 The factors needed to bring the charged bare masses into
 agreement with the neutral bare mass, as in the right-hand 
 panel of Fig.~\ref{pisketch}, are
 \begin{equation}
 Z_{m_d}^{QED} = Z_{m_s}^{QED} = 1.023, \qquad
 Z_{m_u}^{QED}= 1.096 \;. 
 \label{ZDash}
 \end{equation}
 As seen in Fig.~\ref{bareplot} this $Z$ factor depends linearly 
 on the quark charge squared. Hence, we can write
 \begin{equation} 
  \delta \mu^D_q =  (1 + K e_q^2 ) \delta \mu_q
  = (1 + K Q_q^2 e^2 ) \delta \mu_q  \;, 
 \label{lead_Dashen} 
 \end{equation} 
 for some constant $K$. 
 By construction, this simplifies the neutral mesons as they will all
 lie on the same line, see Fig.~\ref{pisketch}.

 \begin{figure}[!htb] 
 \begin{center} 
 \epsfig{file=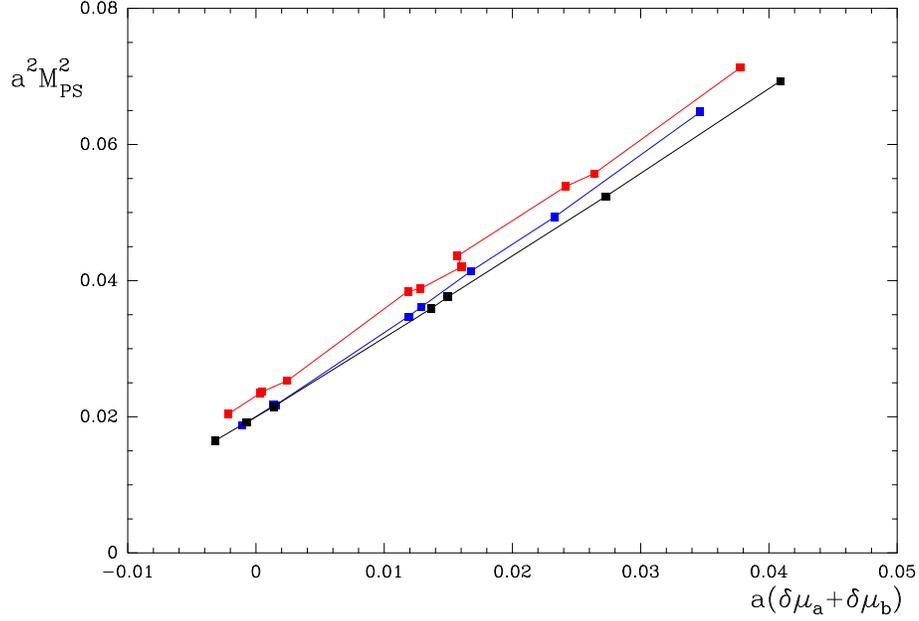, width=12cm} 
 \caption{  Pseudoscalar $M^2_{PS}$ plotted against bare mass
 for the $\pi^+$ (red), $u \bar u$ (blue) and $d \bar d$ (black)
 mesons.  The lines simply connect the points. Error bars are small
 compared with the points.
 Data are from a $32^3 \times 64$ lattice.
 \label{pibare} }
 \end{center} 
 \end{figure} 

 In order to investigate the effect on charged mesons, we
 first consider the $u \bar u, d \bar d$ and $u \bar d\ (\pi^+)$ 
 meson masses plotted as a function of bare quark mass,~Fig.~\ref{pibare}.
 We see that in this plot the two neutral mesons, $u \bar u$ and 
 $d \bar d$, lie on different lines. We also observe that the $\pi^+$
 data do not lie on a smooth curve. This is not due to statistical 
 errors (which are much too small to see in this plot). It is
 because the $\pi^+$ meson mass depends both on $\delta m_u + \delta m_d$, 
 as in pure QCD, but also has a significant dependence on $\dmu - \dmd$, 
 which causes those mesons containing quarks with very unequal masses
 to deviate from the trend. 

 \begin{figure}[!htb]  \begin{center}
 \epsfig{file=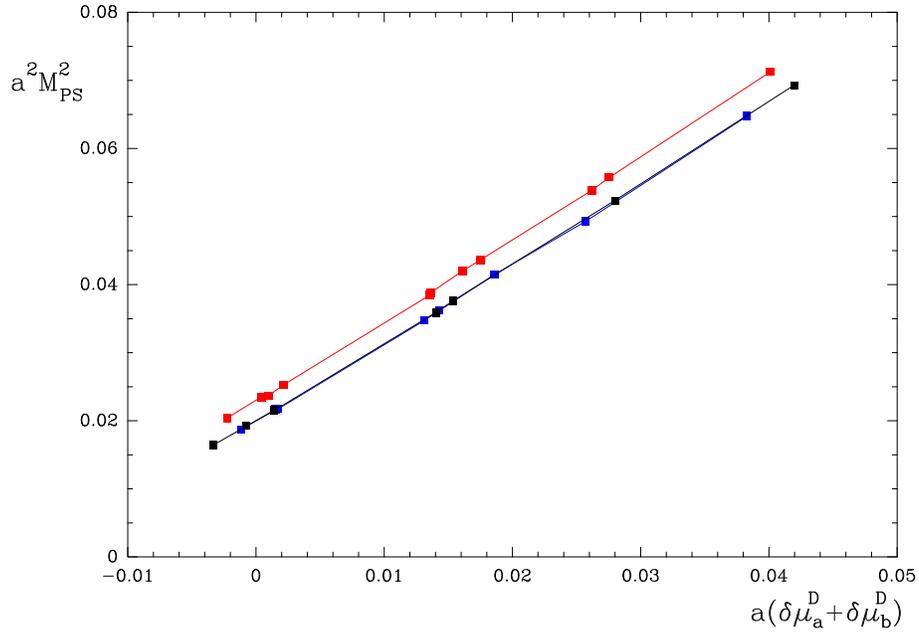, width=12cm} 
  \caption{  The same data as in Fig.~\ref{pibare}, but this time
 plotted against Dashen-scheme quark mass.
 \label{pidash} }
 \end{center}
 \end{figure}

 When we now switch to using the Dashen-scheme quark masses in
 Fig.~\ref{pidash} we see that the graph looks significantly
 different.  The $u \bar u$ and $d \bar d$ mesons now lie on the same
 straight line (this is essentially by construction, since equal
 Dashen-scheme quark mass $\Leftrightarrow$ equal neutral meson
 mass). More interesting is the fact that the ``jiggles" in the
 $\pi^+$ mass are largely removed by plotting against Dashen-scheme
 mass, making it much easier to estimate the EM shift in the $\pi^+$
 mass.

     \section{Dashen scheme quark mass formula \label{Dashq}}

  In order to derive an expression for the meson masses in the
 Dashen-scheme, we start with (\ref{QEDmeson}) and proceed by
 absorbing the QED terms for the neutral pseudoscalar mesons into
 the quark self-energy by making the definition
 \begin{eqnarray}
 \vmq^D &=& \vmq + \Big\{ 
 \half c (\dmu + \dmd + \dms)  + 
  \half\gamma_0^{EM} (e_u^2 \dmu + e_d^2 \dmd + e_s^2 \dms)
  \label{Dashform}\\ &&{} \! \!
 + \gamma_1^{EM} e_q^2 \vmq 
 + \gamma_4^{EM} (e_u^2 +e_d^2 + e_s^2) \vmq
 +  \gamma_5^{EM} e_q (e_u \dmu + e_d \dmd + e_s \dms) \Big\}/\alpha
 \; . \nonumber \end{eqnarray}
 At present we are neglecting $\gamma_0^{EM}$ and $ \gamma_5^{EM}$
 because we are working on a symmetric background, $\delta m_q = 0$,
 and absorbing $\gamma_4^{EM}$ 
 into the coefficient
 $\alpha$ because we only have data at one value of $\beta_{QED}$.
 This means that only the $\gamma_1^{EM}$ term is used in
 calculating $\vma^D$, giving a simple multiplicative transformation
 from bare mass to Dashen scheme mass. Most of the other terms
 in~(\ref{Dashform}) represent off-diagonal terms in the quark mass
 $Z$ matrix. There are many more mixing terms possible in 
 QCD+QED than in pure QCD, but most of them first occur in 
 diagrams with a large number of gluon and quark loops,
 as can be seen in Fig.~\ref{Feyn}, so they are probably
 rather small. 

    Substituting~(\ref{Dashform}) into~(\ref{QEDmeson}) we are left
 with the simpler formula
  \begin{eqnarray}
 M^2(a \bar b) &=& M^2 + \alpha ( \vma^D + \vmb^D)
 + \beta_0 {\ts \frac{1}{6}} ( \dmu^2 + \dmd^2 + \dms^2)
 \label{DashMes} \\ &&{}
 + \beta_1 ( (\vma^D)^2 + (\vmb^D)^2)
 + \beta_2 ( \vma^D - \vmb^D)^2
 + \beta_2^{EM} (e_a - e_b)^2
 \nonumber \\ &&{}
 + \gamma_2^{EM} (e_a - e_b)^2 (\vma^D + \vmb^D )
 + \gamma_3^{EM} (e_a^2 - e_b^2) ( \vma^D - \vmb^D )
 \; . \nonumber \end{eqnarray}
 In~(\ref{DashMes}) all the EM terms vanish for neutral
 mesons ($e_a = e_b$), leaving
\begin{eqnarray}
 M^2_{neut}(a \bar b) &=& M^2 + \alpha ( \vma^D + \vmb^D)
 + \beta_0 {\ts \frac{1}{6}} ( \dmu^2 + \dmd^2 + \dms^2)
 \label{Dashneut} \\ &&{}
 + \beta_1 \left( (\vma^D)^2 + (\vmb^D)^2 \right)
 + \beta_2 \left( \vma^D - \vmb^D \right)^2\; ,
 \nonumber \end{eqnarray}
 which clearly has no references to any $EM$~coefficient,
 or to any charges $e_q$. Hence, by construction, the mass
 of the neutral pseudoscalar mesons comes purely from
 the quark masses, and has no electromagnetic contribution.
 The formula simplifies even further if we consider 
 a flavour-diagonal meson
\begin{equation}
 M^2(a \bar a) =  M^2 + 2 \alpha \vma^D 
 + \beta_0 {\ts \frac{1}{6}} ( \dmu^2 + \dmd^2 + \dms^2)
 + 2 \beta_1 (\vma^D)^2  \;. 
 \label{Dashdiag} 
  \end{equation}
 This agrees with what we see in Figs.~\ref{pisketch} and \ref{pidash},
 with the different flavour-diagonal
 mesons all lying on the same curve when
 plotted against the Dashen quark mass. 

  In the Dashen scheme
   the electromagnetic contribution to the meson mass is 
 \begin{eqnarray}
 M^2_\gamma(a \bar b) &=& \beta_2^{EM} (e_a - e_b)^2
 + \gamma_2^{EM} (e_a - e_b)^2 (\vma^D + \vmb^D )
 \label{MQED}
 \\ &&{}
 + \gamma_3^{EM} (e_a^2 - e_b^2) ( \vma^D - \vmb^D ) \;, 
 \nonumber \end{eqnarray}
 while the QCD contribution is
  \begin{eqnarray}
 M^2_{QCD}(a \bar b) &=& M^2 + \alpha ( \vma^D + \vmb^D)
 + \beta_0 {\ts \frac{1}{6}} ( \dmu^2 + \dmd^2 + \dms^2)
 \label{MQCD} \\ &&{}
 + \beta_1 ( (\vma^D)^2 + (\vmb^D)^2)
 + \beta_2 ( \vma^D - \vmb^D)^2\;. 
 \nonumber \end{eqnarray}


  Dashen's theorems~\cite{Dashen} state that in the limit of an
 exact SU(3) chiral
 symmetry, the neutral mesons have zero electromagnetic self energy; and that
 the charged mesons electromagnetic self-energies are given by a single
 constant. Our formulation is such as to maintain the vanishing
 electromagnetic self-energy of the neutral mesons away from the chiral limit.
 The $\beta_2^{EM}$ term of our expansion is the generalisation of Dashen’s
 result, where, in the absence of any strong SU(3) breaking, the
 electromagnetic self-energy is proportional to the charge-square of the
 meson. The terms involving $\gamma^{EM}$ therefore encode the deviations
 associated with leading-order SU(3) breaking of the strong interaction, as
 anticipated by Dashen.

 \section{Scheme dependence \label{scheme}} 

  We can calculate electromagnetic contributions to the meson masses
 from~(\ref{MQED}) in our scheme, but in order to compare our results
 with those obtained by other groups, we need to be able to quote
 the QED contribution in other schemes, in particular
 $\overline{MS}$. 
 
 To illustrate the issue of scheme dependence, consider the splitting
 between the $K^0$ and $ K^+$ mesons. In the real world the $K^0$-$
 K^+$ splitting comes partly from QED effects, and partly from the
 $m_d, m_u$ mass difference, which we consider to be the QCD part of
 the splitting. The ordering of the physical states, with the $K^0$
 heavier than the $K^+$ suggests that the quark mass effect dominates,
 but we expect that there is still a QED contribution of comparable
 magnitude.

 Naively, one might think that this QED contribution may be easily
 determined by performing a simulation with $m_u = m_d$. In this case,
 there will be no splitting from QCD, so the result will give the
 splitting due to QED alone.
 In pure QCD, setting $m_u = m_d$ is unproblematic as equal bare mass
 implies equal renormalised mass, regardless of scale or scheme.
 However in QED+QCD, mass ratios between quarks of different charges
 are not invariant.  The anomalous dimension of the quark mass now
 depends on the quark charge; at one-loop
 \begin{equation} 
  \gamma_m = 6 C_F g^2 + 6 Q_f^2 e^2 + \cdots 
 \end{equation} 
 so the $u$ mass runs faster than $d$ mass.
  If $m_u = m_d$ in one scheme, this will not be true in another.
 This also implies that there is no good way to compare masses at
 the physical $e^2$ with pure QCD masses at $e^2 = 0$.


 \subsection{Changing Scheme} 

   To calculate the electromagnetic part of the meson mass
 we take the difference between the mass calculated in the 
 full theory, QCD+QED, ($g^2$ and $e^2$ both non-zero)
 and subtract the mass calculated in pure QCD, ($e^2=0$):  
 \begin{equation} 
   M_\gamma^2 = M^2(g^2,e_\star^2,m_u^\star, m_d^\star, m_s^\star)
 - M^2(g^2,0,m^{QCD}_u, m^{QCD}_d,m^{QCD}_s) \;.  
 \end{equation} 
 where $e_\star$ is the physical value of the electromagnetic
 coupling, corresponding to $\alpha_{EM} = 1/137.$
 In the full theory the physical quark masses are well defined: 
 we can fix the three physical quark masses by using three 
 physical particle masses (the $\pi^0, K^0$ and $K^+$ would 
 be a suitable choice). 
 In the full theory we should use the physical quark masses, $m^\star$, 
 but we also have to specify which quark masses we are going to 
 use in the pure QCD case, (which is, after all, an unphysical 
 theory). Different ways of choosing the $m^{QCD}$ will give
 different values for the electromagnetic part of the meson mass. 

   One prescription for choosing the quark masses in the 
 (unphysical) pure QCD case is to use the neutral meson masses.
 We could tune $m^{QCD}$ by requiring   
 \begin{equation} 
 M^2_{q \bar q}(g^2, e_\star^2, m_u^\star, m_d^\star, m_s^\star) 
 = M^2_{q \bar q}(g^2,0, m^{QCD}_u, m^{QCD}_d,m^{QCD}_s)
 \end{equation} 
  Since the QCD+QED mass matches the QCD mass, this scheme has zero
 EM contribution to neutral pseudoscalars by definition.
 This is our Dashen scheme, discussed above. 
 In this scheme, $M^2_\gamma$ is zero for neutral pseudoscalar
 mesons, and is given by the simple formula~(\ref{MQED}) for
 charged mesons.



   A more conventional choice is to choose $m^\star$ and 
 $m^{QCD}$ the same in $\overline{MS}$ at some particular 
 scale.  
In this case, we are now presented with the task of determining the
quark masses in a certain scheme (e.g. the Dashen scheme) given fixed
$\overline{MS}$ masses.
Hence we need to calculate the Dashen quark masses by renormalising
from $\overline{MS}$ to the Dashen scheme:
   \begin{eqnarray}
 m^D(g^2, e_\star^2) &=& Z_m(g^2, e_\star^2, \mu^2)
                              m^{\overline{MS}}(\mu^2) \;, \\
 m^D(g^2, 0, \mu^2) &=&   Z_m(g^2, 0, \mu^2) m^{\overline{MS}}(\mu^2)
 \;.  \nonumber 
 \end{eqnarray} 
 However, since the renormalisation factor $Z_m$ depends on both $g^2$
 and $e^2$, the Dashen mass in pure QCD would not be the same as the
 Dashen mass in the physical QCD+QED theory:
 \begin{equation} 
 m^D_{QCD} \equiv 
 m^D(g^2, 0, \mu^2) = \frac{  Z_m(g^2, 0, \mu^2) }
                          {  Z_m(g^2, e_\star^2, \mu^2) } 
 m^D(g^2, e_\star^2) \equiv Y_m( g^2, e_\star^2, \mu^2) m^D(g^2, e_\star^2)
 \;.
 \label{mDQCD}
 \end{equation} 
 Hence the Dashen mass is rescaled by a
 renormalisation constant ratio which we denote $Y_m$. 
 
 Now, we know in principle what the QCD mass we should subtract
 is, it is the mass we get by substituting $e^2=0, m^D = m^D_{QCD}$  
 into our fit formula. 
So now it is a matter of determining the ratio $Y_m$ in~(\ref{mDQCD})
To proceed, we note that we already know the renormalisation factor
from bare lattice mass to Dashen mass, equation~(\ref{lead_Dashen})
and~(\ref{Dashform}):
 \begin{eqnarray} 
 Y_m^{latt \to D} &=& 1 + \frac{\gamma_1^{EM}}{\alpha} e^2 Q_q^2 
 \label{Ydashlat} 
 \\ &=& 1 + \alpha_{EM} Q_q^2 \; 2.20(9)  \nonumber \;. 
 \end{eqnarray} 
 We also need the renormalisation factor from bare lattice 
 mass to $\overline{MS}$, which can be estimated from lattice 
 perturbation theory~\cite{offshell}. 
 Fortunately, all pure QCD diagrams with only gluons
 and quarks cancel because we are looking at a ratio of $Z$ factors,
 so the leading contribution comes from the 
 1-loop photon diagram, giving 
 \begin{eqnarray} 
 Y_m^{latt \to \overline{MS}} 
 &=& 1 + \frac{e^2 Q_q^2}{16 \pi^2} \left( -6 \ln a \mu + 12.95241 
 \right) \nonumber \\
 &=& 1 + \alpha_{EM} Q_q^2 \; 1.208\;.
 \end{eqnarray} 
 The numerical value in the second line is obtained for $\mu=2$~GeV
 and the value of the lattice spacing in our simulations, $a^{-1} =
 2.9$~GeV (see Table~\ref{physmass}). However, the one-loop result is
 not the full answer, there will be higher order diagrams, with one
 photon plus any number of gluons, giving contributions $\sim e^2 g^2,
 e^2 g^4, \dots$ To account for these unknown terms we add an error
 $\sim \pm 30$\% to the coefficient, giving
 \begin{equation} 
 Y_m^{latt \to \overline{MS}} = 1 + \alpha_{EM} Q_q^2 \; 1.2(4) \;. 
 \end{equation} 
 Combining this with~(\ref{Ydashlat}) gives us the conversion factor
 from the Dashen scheme to $\overline{MS}$ at $\mu=2$~GeV for our
 configurations ($a^{-1} = 2.9$~GeV)
 \begin{equation}
 Y_m^{D \to \overline{MS}} = 1 - \alpha_{EM} Q_q^2 \; 1.0(5) 
 \equiv 1 + \alpha_{EM} Q_q^2 \Upsilon^{D \to \overline{MS}} \;. 
 \label{Yval} 
 \end{equation} 

  We are now ready to write the transformation formula 
 from Dashen scheme $M_\gamma$ to $M_\gamma$ in $\overline{MS}$. 
 In the Dashen scheme
 \begin{equation} 
 \left[M^2_\gamma \right]^D
 = M^2(g^2, e^2, [m_u^\star]^D,  [m_d^\star]^D,  [m_s^\star]^D )-
  M^2(g^2, 0, [m_u^\star]^D,  [m_d^\star]^D,  [m_s^\star]^D )
 \label{MgD} 
 \end{equation} 
 with the same Dashen-scheme quark masses in both terms.
 In $\overline{MS}$ 
 \begin{equation} 
 \left[M^2_\gamma \right]^{\overline{MS}} 
 = M^2(g^2, e^2, [m_u^\star]^D,  [m_d^\star]^D,  [m_s^\star]^D )-
  M^2(g^2, 0, [\tilde m_u]^D,  [\tilde m_d]^D,  [\tilde m_s]^D )
 \label{MgMSb} 
 \end{equation} 
 where $[\tilde m_q]^D$ is given by~(\ref{mDQCD})
 \begin{equation} 
 [\tilde m_q]^D = \left( 1 + \alpha_{EM} Q_q^2 
 \Upsilon^{D \to \overline{MS}} \right)  [m_q^\star]^D \;. 
 \end{equation} 
 
 Taking the difference between~(\ref{MgMSb}) and~(\ref{MgD}) gives 
 \begin{equation} 
 \left[M^2_\gamma \right]^{\overline{MS}}\! - \left[M^2_\gamma \right]^D
 \!= M^2(g^2, 0, [m_u^\star]^D,  [m_d^\star]^D,  [m_s^\star]^D ) - 
  M^2(g^2, 0, [\tilde m_u]^D,  [\tilde m_d]^D,  [\tilde m_s]^D )
 \end{equation} 
 which holds for the electromagnetic contribution to any hadron. 
 If we are specifically interested in pseudoscalar mesons, we
 can use the leading order mass formula 
 $M^2(a \bar b) = \alpha (m_a + m_b)$ 
 to give 
 \begin{eqnarray}
 \left[M^2_\gamma(a \bar b) \right]^{\overline{MS}} &=&
 \left[M^2_\gamma(a \bar b) \right]^D 
 - \alpha_{EM}  \Upsilon^{D \to \overline{MS}} 
 \alpha \left[ Q_a^2 [m_a^\star]^D + Q_b^2 [m_b^\star]^D \right]
 \nonumber \\
 &=& \left[M^2_\gamma(a \bar b) \right]^D 
 - \alpha_{EM}  \Upsilon^{D \to \overline{MS}} 
 \half \left[ Q_a^2 M^2(a \bar a) + Q_b^2 M^2(b \bar b) \right] \;. 
 \label{Upsil} 
 \end{eqnarray} 
 This is a rather simple formula, the only difficulty is that at
 present we only have a rather rough value for the
 constant $\Upsilon$.

 \section{Lattice Results \label{results}} 

  The first question to consider is how close our simulation 
 is to the symmetric line, where $M(u \bar u) = M(d \bar d) 
 = M(s \bar s).$ We find that at the simulation point, 
  $M(u \bar u)$ is about 6\% heavier than the other two 
 mesons, so we are not quite at the desired point. 
 In Table~\ref{ksymtab} we show the $\kappa^{sym}_q$
 values determined on our two large-volume ensembles. 
 In our fits we make a Taylor expansion about the 
 symmetric point of Table~\ref{ksymtab}, not about 
 our simulation point. (The displacement is 
 rather small, the difference is in the fifth significant figure.) 
 
 \begin{table}[htb] \begin{center}
 \begin{tabular}{|c|llc|} 
 \hline 
 flavour & \multicolumn{1}{c}{ $ 32^3 \times 64$} &
 \multicolumn{1}{c}{ $48^3 \times 96$ } & simulation \cr
 \hline 
  $n$  & $0.1208142(14)$ & $0.1208135(9) $&   \cr
 $d,s$ & $0.1217026(5)$ & $0.1217032(3) $& 0.121713  \cr 
 $u$ & $0.1243838(10) $& $0.1243824(6) $& 0.124362  \cr 
 \hline  
 \end{tabular} 
 \caption{ The $\kappa$ values of the symmetric point, 
 determined from fits to the pseudoscalar meson data. 
  \label{ksymtab}} 
 \end{center} 
 \end{table} 

 The next question is whether we have the value of $\mbar$ 
 correctly matched to the physical value. 
 This is checked by comparing the averaged pseudoscalar 
 mass squared, $X^2_\pi$,~(\ref{Xpidef}), with the 
 corresponding baryon scale 
 \begin{equation} 
 X^2_N = \third \left[ (M_N^\star)^2 + (M_\Sigma^\star)^2 
 + (M_\Xi^\star)^2 \right] \;. 
 \label{XNdef} 
 \end{equation} 
 We find $X_N/X_\pi = 2.79(3)$, very close to the correct
 physical value, 2.81, showing that our tuning has found the
 correct $\mbar$ value very successfully.

  \subsection{The splitting of the $\pi^+$ and $\pi^0$ masses.} 

  The first quantity we wish to consider is the mass difference 
 between the $\pi^+$ and $\pi^0$ mesons. Since in this case we are 
 calculating a physically observable mass difference there is no
 scheme dependence in the result.

 First we need to find the $\kappa$ values corresponding to
 the physical quark masses. Since we have three quark
 masses to determine we need three pieces of physical 
 input, we choose the masses of the $\pi^0$ and the two kaons
 \begin{eqnarray}
 M_{\pi^0} &=& 134.977  {\rm \ MeV,} \nonumber\\
 M_{K^0} &=& 497.614 {\rm \ MeV,}\\
 M_{K^+} &=& 493.677  {\rm \ MeV}  \nonumber 
 \end{eqnarray} 
 at $\alpha_{EM} = 1/137$. 
  This determines the physical point given in Table~\ref{physmass}. 
 We see very close  agreement between the lattice scale determined
 on the two lattice volumes. 
 
 \begin{table}[htb] \begin{center}
 \begin{tabular}{|c|cc|} 
 \hline 
 & $ 32^3 \times 64$ &  $48^3 \times 96$ \cr \hline
 $ a \dmu^\star$   &  $-0.00834(8)$ &$ -0.00791(4) $\cr
 $ a \dmd^\star $  &  $-0.00776(7) $ &$ -0.00740(4) $\cr 
 $ a \dms^\star$   & $0.01610(15) $  &$ 0.01531(8) $\cr 
    $ a^{-1}$/GeV   &  2.89(5) &  2.91(3) \cr \hline 
 \end{tabular} 
 \caption{ Bare quark mass parameters at the physical point, 
 and inverse lattice spacing, defined from $X_\pi$. 
 These masses have been tuned to reproduce the real-world 
 $\pi^0, K^0$ and $K^+$ when $\alpha_{EM} = 1/137$. 
 \label{physmass} } 
 \end{center} 
 \end{table} 
 
 Using these quark masses we now have a prediction for the 
 one remaining meson mass, the $\pi^+$. Our values
 on the two lattice spacings are given in Table~\ref{piplus}. 

 \begin{table}[htb] \begin{center}
 \begin{tabular}{|c|ccc|} 
 \hline 
 & $ 32^3 \times 64$ &  $48^3 \times 96$ & Real World \cr \hline
 $M_{\pi^+}$ & 140.3(5) & 139.6(2) & 139.570 \cr 
 $M_{\pi^+} - M_{\pi^0}$ & 5.3(5) & 4.6(2) & 4.594 \cr \hline
 \end{tabular} 
 \caption{ The predicted value of the $\pi^+$ mass, and 
 $\pi^+$-$\pi^0$ splitting, in MeV. 
 \label{piplus}} 
 \end{center} 
 \end{table} 

 \subsection{The $\epsilon$ parameters} 

 The $\pi^+$-$\pi^0$ mass splitting that we presented in the previous
 section is a physically measurable quantity, so it is independent of
 renormalisation.
 However, if we now attempt to divide our hadron masses into a QCD
 part and a QED part, as explained earlier, this is a scheme-dependent
 concept. When we look with greater resolution we see more short
 wavelength photons, which had previously been counted as part of the
 quark mass, and therefore part of the QCD contribution to the mass.

    The traditional way of expressing the electromagnetic
 contributions is through the $\epsilon$ parameters, which 
 measure $M^2_\gamma$ in units of
 \begin{equation}
 \Delta_\pi \equiv M^2_{\pi^+} - M^2_{\pi^0} \;, 
 \end{equation} 
 a natural choice because it is 
 a quantity of a similar origin, and similar order of magnitude. 

  The $\epsilon$ parameters are defined by~\cite{FLAG} 
 \begin{eqnarray}
 M^2_\gamma(\pi^0) = M^2_{\pi^0}(g^2,e^2) - M^2_{\pi^0}(g^2,0) 
 &=& \epsilon_{\pi^0} \Delta_\pi \;, \nonumber \\
 M^2_\gamma(K^0) = 
 M^2_{K^0}(g^2,e^2) - M^2_{K^0}(g^2,0) 
 &=& \epsilon_{K^0} \Delta_\pi \nonumber\;,  \\
 M^2_\gamma(\pi^+) = 
 M^2_{\pi^+}(g^2,e^2) - M^2_{\pi^+}(g^2,0) 
 &=& [1 + \epsilon_{\pi^0} -\epsilon_m] \Delta_\pi\;,  \\
 M^2_\gamma(K^+) = 
 M^2_{K^+}(g^2,e^2) - M^2_{K^+}(g^2,0) 
 &=& \epsilon_{K^+} \Delta_\pi 
 = [1 + \epsilon +\epsilon_{K^0} -\epsilon_m] \Delta_\pi \;. \nonumber
 \end{eqnarray} 
 $\epsilon_{K^+}$ is defined in this way so that the electromagnetic
 contribution to the following quantity has a simple expression
 \begin{equation} 
 [M^2_{K^+} - M^2_{K^0} - M^2_{\pi^+} + M^2_{\pi^0}]_\gamma 
 = \epsilon \Delta_\pi \;. 
 \end{equation} 

   From now on we will neglect the small quantity $\epsilon_m$, 
 the QCD contribution to the $\pi^+$-$\pi^0$ splitting, 
 which comes largely from annihilation diagrams. 
This is a reasonable assumption here since we note that
phenomenological estimates for the this QCD contribution are of order
0.1~MeV (or 2\%) \cite{Gasser:1982ap}, which is within the precision
of our present calculation.

   In the Dashen scheme the $\epsilon$ parameters are simply, 
 \begin{equation} 
 \epsilon_{\pi^0}^D = 0, \qquad \epsilon_{K^0}^D=0, \qquad
 \qquad \epsilon_{\pi^+}^D = 1 \;, 
 \end{equation} 
with the only non-trivial quantity, $\epsilon^D$, given by
 \begin{equation} 
 \epsilon^D = \frac{M_\gamma^2(K^+)}{M_\gamma^2(\pi^+)} -1 
 = \epsilon_{K^+}^D -1
 \end{equation} 
On our two ensembles we find 
 \begin{eqnarray} 
 \epsilon^D = 0.38(10) & \qquad & 32^3 \times 64 \;,\nonumber \\
 \epsilon^D = 0.49(5) && 48^3 \times 96  \;, 
 \end{eqnarray} 
 which agree within errors. In what follows, we use the
 $48^3 \times 96$ value in our calculations. 

    Using (\ref{Upsil}) to transform these numbers into 
 $\overline{MS}$ with the scale $\mu=2$~GeV, we find: 
 \begin{eqnarray} 
 \epsilon_{\pi^0} &=& - \alpha_{EM} \Upsilon^{D \to \overline{MS}} 
 \half \left[\ts  \frac{4}{9} M^2(u \bar u) + \frac{1}{9} M^2(d \bar d)
 \right] /\Delta_\pi = 0.03 \pm 0.02 \;, \nonumber \\
 \epsilon_{\pi^+} &=& 
 \epsilon_{\pi^+}^D - \alpha_{EM} \Upsilon^{D \to \overline{MS}} 
 \half \left[ \ts \frac{4}{9} M^2(u \bar u) + \frac{1}{9} M^2(d \bar d)
 \right] /\Delta_\pi = 1.03 \pm 0.02 \;, \nonumber \\
 \epsilon_{K^0} &=& - \alpha_{EM} \Upsilon^{D \to \overline{MS}} 
 \half \left[\ts  \frac{1}{9} M^2(d \bar d) + \frac{1}{9} M^2(s \bar s)
 \right] /\Delta_\pi = 0.2 \pm 0.1 \;,  \\
 \epsilon_{K^+} &=& \epsilon_{K^+}^D 
                    - \alpha_{EM} \Upsilon^{D \to \overline{MS}} 
 \half \left[ \ts \frac{4}{9} M^2(u \bar u) + \frac{1}{9} M^2(s \bar s)
 \right] /\Delta_\pi = 1.7 \pm 0.1 \;, \nonumber \\ 
 \epsilon &=& \epsilon^D 
                    - \alpha_{EM} \Upsilon^{D \to \overline{MS}} 
 \half \left[\ts  \frac{4}{9} M^2(u \bar u) - \frac{1}{9} M^2(d \bar d)
 \right] /\Delta_\pi =  0.50 \pm 0.06\;. \nonumber 
 \end{eqnarray} 
 In all cases we are resolving more photons in $\overline{MS}$, 
 and so converting some fraction of the quark mass into electromagnetic
 energy. This has very little effect in the pions because both 
 quarks are very light, but a much larger effect in the kaons
 because the strange quark is heavier, and the photon cloud has 
 a mass proportional to the quark mass. 

%
%
%

%
 
 \section{Conclusions} 

 We have investigated isospin breaking in the pseudoscalar
 meson sector from lattice calculations of QCD+QED. This allows
 us to look simultaneously at both sources of isospin breaking, 
 the quark mass differences, and the electromagnetic interaction, 
 which are of comparable importance. 

    The physical mass differences between the different particles
 are directly observable, and so must be independent of the 
 renormalisation scheme and scale used. When we try to go beyond
 this, to say what fraction of a hadron's mass-squared comes from 
 QCD, and from QED, this no longer holds --- changing our resolution
 changes the fraction. We understand this effect, both formally, in terms
 of the dependence of the mass renormalisation constant on the 
 electromagnetic coupling, and physically, in terms of the 
 quark mass gaining a contribution from its associated photon cloud. 

    With this understanding, we calculate the electromagnetic 
 contributions to hadron masses in the Dashen scheme, 
 which is easy to implement on the lattice, and then convert 
 these values into the more conventional $\overline{MS}$ scheme. 

   We are also investigating the isospin violating mass splittings
 in the baryon sector~\cite{Horsley:2015}, as well as the 
 decomposition of these mass differences into QCD and QED parts, 
 both in the Dashen scheme, and in $\overline{MS}$.


 \section*{Acknowledgements}
 
 The computation for this project has been carried out on 
 the IBM BlueGene/Q using DIRAC 2 resources (EPCC, Edinburgh, UK),
 the BlueGene/P and Q at NIC (J\"ulich, Germany), the
 SGI ICE 8200 and Cray XC30 at HLRN (The North-German Supercomputer
 Alliance) and on the NCI National Facility in Canberra, Australia
 (supported by the Australian Commonwealth Government).
     
 Configurations were generated with a new version of the 
 BQCD lattice program~\cite{nakamura10a}, modified to
 include QCD+QED. Configurations were analysed using 
 the Chroma software library~\cite{edwards04a}. 

 HP was supported by DFG Grant No. SCHI~422/10-1.
 PELR was supported in part by the STFC under contract ST/G00062X/1,
 RDY was supported by the Australian Research Council Grant
 No. FT120100821 and DP140103067
 and JMZ was supported by the Australian Research Council Grant
 No. FT100100005 and DP140103067. We thank all funding agencies.


 \end{document}